# Theoretical investigation of electronic specific heat of two-band superconductors

E.G.Maksimov[1], A.E.Karakozov[2], B.P.Gorshunov[3,4], E.S.Zhukova[3,4], Ya. G. Ponomarev[5], M.Dressel[6]

[1] P.N. Lebedev Physical Institute, Russian Academy of Sciences, 119991 Moscow, Russia

[2] L.F. Vereshchagin Institute of High Pressure Physics, Russian Academy of Sciences, Troitsk, Moscow region, 142190 Russia

[3] A.M. Prokhorov Institute of General Physics, Russian Academy of Sciences, 119991 Moscow, Russia

[4] Moscow Institute of Physics and Technology (State University), 141700, Dolgoprudny, Moscow Region, Russia

[5] Faculty of Physics, M.V. Lomonosov Moscow State University, 119991 Moscow, Russia

[6] 1. Physikalisches Institut, Universität Stuttgart, Stuttgart, Germany

**Abstract**

We apply the generalized two-band model of superconductivity to calculate the electronic contribution to the specific heat, superconducting gaps and electron-boson coupling constants of the multi-band layered superconductors $MgB_2$ and $Ba(Fe_{0.925}Co_{0.075})_2As_2$. For both compounds, the obtained temperature-behavior of the specific heat well describes the experimental data. We compare our findings with the "two-band $\alpha$ - model" that is frequently used to calculate the electronic specific heat, and that is applicable only for superconductors with strong interband scattering ("dirty" enough superconductors).

**1. Introduction.**

After the discovery of layered superconductors – magnesium diboride and iron-pnictides/chalcogenides – interest has re-emerged to examine theoretically the electronic specific heat $C_s$ of two-band superconductors (for a review see Ref. [1]). In superconducting $MgB_2$ and $Ba(Fe_{1-x}Co_x)_2As_2$ the two-gap superconductivity reveals itself in a clearly anomalous temperature behavior of the specific heat, $C_s(T)$ [2,3]. The anomaly appears in the form of characteristic features at relatively low temperatures $T^* << T_c$, where bumps in $C_s(T)$ are smeared by interband interaction and anomalies are seen corresponding to the specific heat of the lower-gap ($\Delta_{min}$) superconducting (SC) band. In case of *independent* bands these jumps would be "real" specific heat jumps at corresponding critical temperatures $T^* \sim T_c^{(0)} \approx \Delta(0)/1.764$. Specific manifestations of two-band superconductivity are also found in *ab initio* calculations of the electron-phonon interaction (EPI) in MgB$_2$ single crystals [4]. However, it is not straightforward to use these results in order to analyze *real* samples with impurities because impurity scattering unavoidably leads to renormalization of the intraband as well as interband EPI constants. To analyze the $C_s(T)$ behaviors, an empirical technique is frequently used known as a "two-band $\alpha$-model" [2]. Within this model each band is treated separately [5] and interband interaction is completely ignored. This way it misses the most important property that has to be taken into account when considering the two-band system.

In this communication we present the results of a theoretical study of specific heat of a two-band superconductor with the interband interaction taken into account. While calculating $C_s(T)$, we apply the $\alpha$-model-like strong-coupling generalization of the two-band Bardeen-Cooper-Schrieffer (BCS) equations that has been used by us earlier to analyze the tunnel experimental data of Mg$_{1-x}$Al$_x$B$_2$ [6] and the optical experiments on Ba(Fe$_{0.9}$Co$_{0.1}$)$_2$As$_2$ [7].

## 2. General properties of two-band superconductors

The two-band BCS equations that determine the temperature dependence of the SC gaps can be written as [8]:

$$\Delta_I(T) = \sum_{J=1,2} \lambda_{IJ} \Delta_J(T) \int_0^{\Omega_c} d\xi \tanh\left(E_J(\xi)/2T\right)/E_J(\xi) \quad (1),$$

$$E_J(\xi) = \sqrt{\xi^2 + \Delta_J^2(T)} \quad . \quad (2).$$

Here $\Omega_{\tilde{n}}$ is the characteristic phonon frequency and $E_J(\xi)$ is the quasiparticles spectrum in the band J (J = 1, 2). The renormalized EPI coupling constants are expressed via BCS paring constants $\tilde{\lambda}_{IJ}^0$ as

$$\tilde{\lambda}_{IJ}^0 = \lambda_{IJ}^0 - \mu_{IJ}^* \quad , \quad (3),$$

where $\lambda_{IJ}^0$ (I = 1, 2 and J = 1, 2) are the EPI coupling constants and $\mu_{IJ}^*$ are the Coulomb pseudopotentials

$$\lambda_{IJ} = \tilde{\lambda}_{IJ}^0 / (1 + \lambda_{II}^0 + \lambda_{I \neq J}^0) \quad . \quad (4).$$

It is convenient to rewrite the system (1) in the form [6]

$$\int_0^{\Omega_{\tilde{n}}} d\xi \tanh[E_1(\xi)/2T]/E_1(\xi) = \tilde{\lambda}_{22} - \tilde{\lambda}_{12}\theta(T) \quad (5),$$

$$\int_0^{\Omega_{\tilde{n}}} d\xi \tanh[E_2(\xi)/2T]/E_2(\xi) = (\tilde{\lambda}_{11} - \tilde{\lambda}_{21})/\theta(T) \quad (6),$$

$$\theta(T)\ln\theta(T) = -[\tilde{\lambda}_{11} - \tilde{\lambda}_{22} - \delta n_{12}(T)]\theta(T) - \tilde{\lambda}_{12}\theta^2(T) + \tilde{\lambda}_{21} \quad , \quad (7),$$

where $\theta(T) = \Delta_2(T)/\Delta_1(T)$ is the gap ratio, $\delta n_{12}(T) = n_1(T) - n_2(T)$ and $n_J(T)$ is the normalized quasi-particles number for the band J:

$$n_J(T) = 2\int_0^\infty d\xi \, f[E_J(\xi)/T]/E_J(\xi) \tag{8}$$

with $f[E_J(\xi)/T]$ being the quasi-particles distribution function and $\tilde{\lambda}_{IJ}$ the effective coupling constants:

$$\tilde{\lambda}_{IJ} = \frac{\lambda_{IJ}}{\lambda_{11}\lambda_{22} - \lambda_{12}\lambda_{21}} = \frac{\lambda_{IJ}}{\lambda_{11}\lambda_{22}} \frac{1}{[1 - \lambda_{12}\lambda_{21}/(\lambda_{11}\lambda_{22})]} \tag{9}.$$

To obtain a self-consistent solution, any convenient pair of equations of (5-7) can be taken. Obviously, the effective coupling constants $\tilde{\lambda}_{IJ}$ given in Eq. (9), especially the *interband* constants (I ≠ J), are strongly sensitive to the relative magnitude of the interband interaction. (Typical values of $\tilde{\lambda}_{IJ}$ are given in Table I). The situation that the determinant of the system (1) vanishes, $\lambda_{11}\lambda_{22} - \lambda_{12}\lambda_{21} \to 0$, does not represent any special interest; in this case the gap ratio does not depend on temperature, $\theta(T) = \lambda_{21}/\lambda_{11} = \lambda_{22}/\lambda_{12} = const$, as can easily be shown using (1).

In the following, using the available data on the EPI in layered superconductors [4], we consider only positive effective coupling constants $\tilde{\lambda}_{IJ}$ (9).

For T=0 the task is reduced to find a suitable $\theta(0)$ that solves Eq. (7)

$$\tilde{\lambda}_{11} - \tilde{\lambda}_{22} = -\ln\theta(0) - \tilde{\lambda}_{12}\theta(0) + \tilde{\lambda}_{21}/\theta(0) \tag{10}$$

with subsequent determination of the gaps $\Delta_{1,2}(0)$ according to the following expressions

$$\ln[2\Omega_c/\Delta_1(0)] = \tilde{\lambda}_{22} - \tilde{\lambda}_{12}\theta(0), \quad \ln[2\Omega_c/\Delta_2(0)] = \tilde{\lambda}_{11} - \tilde{\lambda}_{21}/\theta(0) \tag{11}.$$

Let us consider some general properties of the two-band superconductor that is described by equations (5,6,7), at finite temperatures and under the assumption $\Delta_1(0) > \Delta_2(0)$. It is easy to show that in this case, due to different amount of quasiparticles (8) in the bands, $n_1(t) < n_2(t)$, and when the temperature increases, the gap ratio $\theta(T)$ will decrease down to its minimal value reached at $T_c$

$$\theta(T_c) = \frac{1}{2\tilde{\lambda}_{12}}\left[-\tilde{\lambda}_{11} + \tilde{\lambda}_{22} + \sqrt{(\tilde{\lambda}_{11} - \tilde{\lambda}_{22})^2 + 4\tilde{\lambda}_{12}\tilde{\lambda}_{21}}\right] \tag{12}$$

with the critical temperature determined from

$$\ln\frac{2\gamma_E\Omega_{\tilde{n}}}{\pi T_c} = \frac{1}{2}\left[\tilde{\lambda}_{11} + \tilde{\lambda}_{22} - \sqrt{(\tilde{\lambda}_{11} - \tilde{\lambda}_{22})^2 + 4\tilde{\lambda}_{12}\tilde{\lambda}_{21}}\right] \tag{13}.$$

We note that in the above two expressions the signs of the square roots correspond to the case $\tilde{\lambda}_{IJ} > 0$, otherwise these signs should be inverted. Expressions (10-13) allow determining important parameters of the theory – the ratios of the superconducting gaps to the critical temperature $\alpha_J = \Delta_J(0)/T_c$:

$$\alpha_1 = \alpha_0 e^{\Lambda_1}, \quad \alpha_2 = \alpha_1 \theta(0) \tag{14},$$

$$\Lambda_1 = \tilde{\lambda}_{12}\theta(0) - \frac{1}{2}\left[\sqrt{(\tilde{\Lambda}_{12})^2 + 4\tilde{\lambda}_{12}\tilde{\lambda}_{21}} - \tilde{\Lambda}_{12}\right] \tag{15},$$

where $\alpha_0 = \pi/\gamma_E \approx 1.764$ and $\tilde{\Lambda}_{12}$ is the right part of expression (10). The so determined parameters $\Lambda_{1,2}$ (and $\alpha_J$) depend only on the values of $\theta(0)$ and of effective *interband* constants $\tilde{\lambda}_{IJ}$. The three parameters $\theta(0)$, $\tilde{\lambda}_{12}$ and $\tilde{\lambda}_{21}$ represent a complete set of the two-band BCS equations for the reduced gaps $\delta_J(t) = \Delta_J(t)/\Delta_J(0)$ at finite temperatures:

$$\ln \delta_1(t) = -n_1(t) - \tilde{\lambda}_{12}\theta(0)\{1 - \delta_2(t)/\delta_1(t)\} \tag{16},$$

$$\ln \delta_2(t) = -n_2(t) - \tilde{\lambda}_{21}/\theta(0)\{1 - \delta_1(t)/\delta_2(t)\} \tag{17},$$

$$n_J(t) = 2\int_0^\infty d\omega\, f[\alpha_J \varepsilon_J(\omega)/t]/\varepsilon_J(\omega) \tag{18},$$

$$\varepsilon_J(\omega) = \sqrt{\omega^2 + \delta_J^2(t)} \tag{19}.$$

In Eqs. (18) and (19), $\varepsilon_J(\omega)$ is the reduced spectrum of a superconductor and $t = T/T_c$ is the reduced temperature. The quantities $\alpha_{1,2}$, $\tilde{\lambda}_{12}$ and $\tilde{\lambda}_{21}$ that enter Eqs. (16) and (17) characterize the intraband and the interband fluctuations, respectively.

Let us consider the contribution of interband fluctuations. The ratio $\delta_2(t)/\delta_1(t) = \theta(t)/\theta(0) < 1$ and decreases with increasing temperature. For that reason the contribution of interband fluctuations leads to the decrease of the gap $\delta_1(t)$ and to an increase of $\delta_2(t)$; but this contribution will never overcome the intraband one. This implies that for large interband constants the gaps values become equal: $\delta_2(t) \to \delta_1(t)$. Along with that, for $\lambda_{21} \gg \lambda_{12}$ the values of both reduced gaps approach $\delta_0(t)$ - the solution of a standard BCS equation. In such case the amplitudes of reduced gaps, $\delta_{1,2}(t)$, at any temperature do not exceed the value $\delta_0(t)$, and their temperature dependences at $t \to 1$ are weaker compared with the dependence $\delta_0(t)$.

In order to evaluate $\alpha_1$ Eq. (14) can be rewritten in a more visual form

$$\Lambda_1 = \tilde{\lambda}_{12}\{\theta(0) - \theta(T_c)\} > 0 \tag{20}.$$

It is easily seen that even in conditions of a relatively weak interband interaction when the parameter $\theta(0)$ depends mainly on intraband coupling constants and $\theta(T_c) \sim \lambda_{21}$, for $\lambda_{12} > \lambda_{21}$ the ratio $\alpha_1$ of $\Delta_1(0)$ to the critical temperature can be rather large – even for superconductors with weak coupling. (See Table I). Fig. 1 shows the dependence of the reduced gap $\delta_2(t)$ on interband constants values $\tilde{\lambda}_{IJ}$; the data in this figure are calculated for two kinds of superconductors, (a) and (b), that have the same ratios of the gaps $\theta(0)$ and the same ratios $\alpha_{1,2}$ of the gaps to the critical temperature, *but different interband interactions:* (a): $\tilde{\lambda}_{12} = 0.05, \tilde{\lambda}_{21} = 0.1$ ; (b): $\tilde{\lambda}_{12} = 0.22, \tilde{\lambda}_{21} = 0.8$. For comparison, also the reduced gap is plotted according to the two-band $\alpha$-model with $\delta_{1,2}(t)$ taken equal to $\delta_0(t)$.

The temperature dependence of the specific heat $C_s$ of a superconductor is usually characterized by the ratio $\tilde{N}_s(T)/\gamma_n T_c$. In case of a two-band superconductor, the specific heat can be calculated in a usual manner [1,5]. Writing down the expression for the entropy $S_J(T)$ [or normalized entropy $s_J(t) \equiv S_J(t)/\gamma_J T_c$] of a free Fermi gas with the spectrum given by Eqs. (2) and (19)

$$S_J(T) = 2N_J(0) \int_{-\infty}^{\infty} d\xi \{[E_J(\xi)/T] f(E_J(\xi)) - \ln[f(-E_J(\xi))]\} \qquad (21),$$

$$s_J(t) = 3\alpha_J/\pi^2 \int_{-\infty}^{\infty} d\omega \{[\alpha_J \varepsilon_J(\omega)/t] f(\alpha_J \varepsilon_J(\omega)/t) - \ln[f(-\alpha_J \varepsilon_J(\omega)/t)]\} \qquad (22),$$

where $\gamma_J = 2\pi^2 N(0)/3$ is the coefficient of normal electronic specific heat of the band J ($\gamma_n = \gamma_1 + \gamma_2$), we get

$$\left[\tilde{N}_s(t)/\gamma_n T_c\right]/t = \frac{d}{dt}\left[\frac{\gamma_1}{\gamma_n} s_1(t) + \frac{\gamma_2}{\gamma_n} s_2(t)\right]. \qquad (23).$$

The dependences of the electronic specific heat on interband interaction, again for the two kinds of superconductors, (a) and (b), are shown in Figs. 2 and 3. For comparison, these figures contain also the correspondent curves calculated within the two-band $\alpha$-model [2]. Fig. 2 clearly demonstrates all peculiarities of the temperature behavior of the specific heat of a two-band superconductor. There is a feature around $T^*(\alpha_2)$ - a smeared, by the interaction $\tilde{\lambda}_{21}$, kink in the specific heat of the band with a smaller gap. In the limit $\tilde{\lambda}_{21} \gg 1$ [corresponding to $\delta_2(t) \to \delta_0(t)$, as we have seen, i.e., to the two-band $\alpha$-model] the kink becomes much smoother. (In the same manner the interaction $\tilde{\lambda}_{12}$ smoothens the kink around $T_c$ that is connected to the first band). From Eq. (23) it is seen that the feature originating from the smaller gap band is more developed when the weight $\gamma_2/\gamma_n$ of a separate band is large enough and when $\theta(0) \ll 1$ (at $T^* \ll T_c$).

We conclude this section by briefly discussing how the two-band $\alpha$-approach can be used for determination of the gaps values from the temperature dependences of the specific heat. One of the merits of the two-band $\alpha$ model is the possibility to correctly determine the value of the smaller gap ($\alpha_2$) by considering the lower-temperature interval of the experimental data; after that the larger gap ($\alpha_1$) can also be found according to the temperature interval closer to $T_c$ – from to the jump of the specific heat. The amplitude of this jump is easily found within the two-band $\alpha$-model by direct calculations of the normalized specific heat (22) $\tilde{C}_s$ at the transition point, using the known dependence $\delta_0(t \to 1)$ from the BCS theory [9]

$$\tilde{N}_{sJ} = 1 + \frac{12}{7\zeta(3)} \left(\frac{\alpha_J}{\alpha_0}\right)^2 \tag{24},$$

$$\tilde{N}_s = 1 + 1.426 \left(\frac{\alpha_1}{\alpha_0}\right)^2 \frac{\gamma_1 + \gamma_2 \theta^2(0)}{\gamma_n} ; \tag{25}.$$

Here $\zeta$ is the zeta-function. According to Eq. (16) and (17), the two-band $\alpha$-approach is, indeed, "theoretically" correctly describing the temperature dependences of the gaps values (and of the specific heat) at temperatures $T \ll T^*$ (see also Figs. 2,3 where the symbols represent the accuracy that could be obtained in experiment). However, in order to approximate, within the two-band $\alpha$-model, the experimental curve in a larger temperature interval up to $T \sim T^*$, one would have to shift the magnitude of the smaller gap ($\alpha_2$) relative to its real value. Note, however, that even this kind of fitting the data can be realized only in case of relatively strong interband interaction (or in "dirty" enough superconductors). Due to a weaker temperature dependence of $\delta_{1,2}$ compared to $\delta_0(t)$, the specific heat jump in the two-band BCS-like superconductor appears to be smaller than the jump obtained within the two-band $\alpha$-model. This means that in order to approximate the curve $C_s(t)$ at temperatures around $T_c$ with the two-band $\alpha$-model, the larger gap has to be reduced relative to its real value. In the two-band $\alpha$-model, the gaps values can be determined more precisely only for interaction strengths $\tilde{\lambda}_{21} \gg 1, \tilde{\lambda}_{12} \ll 1$. Note also, that in contrast to regular superconductors, the values of $\alpha_1 > 1.764$ and of the specific heat jump $\Delta C_s > 1.426$ in the two-band superconductors are not unambiguous indications of a strong EPI.

### 3. Electronic specific heat of layered superconductors

The properties of conventional superconductors at low temperatures are well described within the BCS approach where the EPI coupling constants and the energy $\Omega_{\tilde{n}}$ are determined by the EPI spectral function $\alpha^2(\omega)F(\omega)$. (A detailed discussion is presented in [10]). However, at finite temperatures the properties of real superconductors can be noticeably different from predictions of the BCS theory. In particular, applying the BCS formalism to the strong coupling superconductors overestimates the value of $T_c$ (and, correspondingly, underestimates the value of $\alpha$), see Table I. Nevertheless, while calculating the properties of conventional strong-coupling superconductors the BCS theory can be successfully applied also for finite temperatures basing on empirical recipe known as $\alpha$-model [5]. In particular, to calculate the electronic specific heat within the $\alpha$-model, one has to use the BCS expression (22) with the spectrum $\varepsilon(\omega) = \sqrt{\omega^2 + \delta_0^2 \left(t = T/T_c^{\exp}\right)}$ and to set $\alpha = \alpha^{\exp}$.

The success of the empirical $\alpha$-model is based on the fact that its results can be obtained theoretically by formal introduction into the BCS equation kernel of the correction term $k_\alpha = T_c^{\exp}/T_c^0$

$$\ln[\Delta(T)/\Delta(0)] = -2\int_0^{\Omega_c} d\xi\, f\left\{\left(T_c^{\exp}/T_c^0\right)\left(E_J(\xi)/T\right)\right\}/E_J(\xi) \qquad (26).$$

This guarantees that the correct value of $T_c = T_c^{\exp}$ is obtained. This procedure fully preserves the form of the BCS equation for the reduced gap $\delta(t)$ as a function of the reduced temperature $t = T/T_c^{\exp}$. To have a clear distinction to the empirical $\alpha$-model, such approach should more appropriately be named as $\alpha$-approximation.

In the two-band case, the $\alpha$-approximation (26) also provides a correct value for $T_c$ keeping the form of Eqs. (16) and (17) unchanged. Such approach, however, is not universal because the strong coupling corrections are different for the two bands. This can be easily shown taking as an example nearly independent ($\tilde{\lambda}_{12} \to 0$, $\tilde{\lambda}_{21} \to 0$) bands with strong coupling $\Delta_1(0) \gg \Delta_2(0)$. In such case $\alpha_1 \to \alpha_0$ and $\alpha_2 \to \alpha_0(T_{c2}^0/T_{c1}^0)$, implying that for reduced temperatures $t \geq t^* = T_{c2}^0/T_{c1}^0 < 1$ the gap value $\delta_2(t) \to 0$. Thus, for finite temperatures the critical temperatures $T_{c1}$ and $T_{c2} \approx T^*$ in independent bands appear to be equal to $T_{c1} = T_c^{\exp}$ and $T_{c2} = T_{c2}^0(T_c^{\exp}/T_{c1}^0) \ll T_{c2}^0$, respectively. This example shows that in the $\alpha$-approximation-like equations (1) the two correction coefficients of the type $k_{1,2} = \beta_{1,2}(\lambda_{IJ})k_\alpha$ have to be taken into account, providing, on the one hand, the correct values of $T_c = T_c^{\exp}$ and, on the other hand, the correct behavior of the weak-coupling band in the limit of independent bands. In Eqs. (16), (17), which describe the reduced gaps, this will lead to the *formal* redefinition of $\alpha_{1,2}$, i.e. of $\Delta_{1,2}(0)$ and $\theta(0)$, since $T_c$ [Eq. (13)] depends only on the coupling constants and is not changed during such redefinition. In the general case, this kind of procedure is difficult to realize. At the same time, for the bands with strong and weak coupling in the limit $\tilde{\lambda}_{12} \ll 1$ the correction coefficients are easily found: $k_1 = k_\alpha$, $k_2 = 1$. Here $\alpha_1 = \alpha_0$, $\alpha_2 = \alpha_2^{\exp}$ and, assuming independent bands, the critical temperatures become $T_{c1} = T_c^{\exp}$ and $T_{c2} = T_{c2}^0$.

The above example can easily be generalized for finite $\tilde{\lambda}_{12}$ by redefining the parameters in Eqs. (16) and (17): $\alpha_2 = \alpha_2^{\exp}$ and $\alpha_1 \to \alpha'$, where $\alpha'$ is determined by Eq. (14) and by a self-consistency condition $\alpha_1'\theta'(0) = \alpha_2^{\exp}$. Here the parameter $t$ keeps the meaning of a reduced temperature [$\delta_{1,2}(t=1) = 0$, $t = T/T_c^{\exp}$]. The auxiliary parameters $\alpha'$ and $\theta'(0)$ are purely formal and carry no particular physics. (We note that the equations within the $\alpha$- and $\alpha'$- approximations keep the form of the BCS equations whose properties are described in details in Section 2).

In Ref. [6], we have applied the $\alpha'$-approximation to theoretically analyze the tunnel experiments on the two-gap superconductors $Mg_{1-x}Al_xB_2$. The EPI "seed" coupling constants have been unambiguously calculated for $\Omega_{\tilde{n}}$ =67.76 meV and for $\mu_{IJ}^* \approx 0.12$ based on the experimental values of $\Delta_{1,2}(0)$ and of Leggett mode energy $\Omega_L(0)$ [11]; note that the Leggett mode energy is extremely sensitive to the interband as well as intraband interactions. Calculations within the $\alpha'$-approximation give a very good agreement with experimental results presented in [12] and obtained not only for the temperature variation of both gaps $\Delta_{1,2}(T)$ but also for the Leggett mode energy $\Omega_L(T)$ (see Fig.4). We have applied the same technique for calculations of the temperature variation of the superconducting condensate density and of the London penetration depth determined for $Ba(Fe_{0.9}Co_{0.1})_2As_2$ in Ref. [13]. The results of our calculations according to the $\alpha$- and $\alpha'$-approximations are presented in Figs. 4 and 5, together with the results obtained according to the two-band $\alpha$-model. We see that the experimental data are well reproduced by calculations within the $\alpha'$–approximations. Unfortunately, there are no specific heat data for the samples measured in Ref. [12,13] and a comparison with our calculations is not possible.

Up to now, for most layered superconductors the properties are not studied in all details, and rather often their characteristics are determined by using indirect methods based on theoretical models. For example, in Ref. [2,3] the two-band $\alpha$-model was employed to determine the gap values $\Delta_{1,2}(0)$ from the temperature variation of the specific heat $C_s(T)$. In these publications a good agreement has been found in the whole temperature interval between the theoretical two-band $\alpha$- calculations and experimental dependences $C_s(T)$ for $Mg^{11}B_2$ ($T_c$=38.7 K) [2] and $Ba(Fe_{0.95}Co_{0.075})_2As_2$ ($T_c$ =21.4 K) [3]. We have used the $\alpha'$-approximation to calculate the specific heat of $Mg^{11}B_2$ and of $Ba(Fe_{0.95}Co_{0.075})_2As_2$ (referred to as Ba(FeCo)As in Table II). The results are presented by Figs. 6 and 7 and in Tables II and III. It can be seen that within the experimental accuracy the specific heat data obtained in Ref. [2,3] are very well approximated by a set of curves with effective coupling constants $\tilde{\lambda}_{12} \leq \tilde{\lambda}_{12}^{max}$, $\tilde{\lambda}_{21} \geq \tilde{\lambda}_{21}^{min}$ and parameters $\alpha_{1,2}$ that are changing from the values indicated in Table II (fit max-min) to the values calculated within the two-band $\alpha$-model. These latter values are obtained for the interband interaction $\tilde{\lambda}_{12} \sim 0.1 << \tilde{\lambda}_{21} \sim 3.5$ (fit min-max).

A comparison of the gaps $\Delta_{1,2}(0)$ in $Mg^{11}B_2$ ($T_c$ =38.7 K) given in Table II with their values in $MgB_2$ ($T_c$ =40.5 K) $\Delta_1(0) \approx 10$ meV, $\Delta_2(0) \approx 2.6$ meV [12] indicating different degree of imperfection of the samples studied in [2] and in [12]. The difference, however, cannot be accounted for by elastic interband scattering on impurities that would lead to an increase of the smaller gap [14]. It looks more probable that the difference should be connected to the different electron-phonon scattering caused by defects in the crystal structure. The smaller gap in $Ba(Fe_{0.925}Co_{0.075})_2As_2$ ($T_c$ =21.4 K) agrees well with the values reliably determined from the optical experiments [13] Ba(Fe$_{0.9}$Co$_{0.1}$)$_2$As$_2$ ($T_c \approx 20$ K): $\Delta_2(0) \approx 15$ cm$^{-1}$ (1.85 meV) and $\Delta_1(0) \approx 30$ cm$^{-1}$ (3.9 meV).

The method we have used for our calculations allows determining not only the effective interband coupling constants, but also all EPI coupling constants of the samples. Assuming $\Omega_{\tilde{n}} = 67.76$ meV (Mg$^{11}$B$_2$), $\Omega_c \approx 30$ meV [Ba(FeCo)As] [15], $\mu^* = 0.12$ and using Eqs. (11), (9), and (4) we find the EPI "seed" constants $\lambda_{IJ}^0$ (see Table III). In Table III we also present, for comparison, the data for MgB$_2$ ($T_c$=40.5 K) determined in our early tunnel experiments [6]. Comparison of the electron-phonon interaction strengths in Mg$^{11}$B$_2$ (data from [2]) and in MgB$_2$ (data from [6]), is really revealing significant differences. First, the interband EPI is much larger in Mg$^{11}$B$_2$ [2] ($\tilde{\lambda}_{12}^0 = 0.085$, $\tilde{\lambda}_{12}^0 = 0.089$) compared to MgB$_2$ [6] ($\lambda_{12}^0 - \mu^* = 0.004$, $\lambda_{21}^0 - \mu^* = 0.037$, see Table III). And, second, an appreciable increase (actually, an equalization) of interband and intraband scattering in the $\pi$-band is observed in Mg$^{11}$B$_2$ [2]: $\tilde{\lambda}_{12}^0 \approx \tilde{\lambda}_{12}^0$. This kind of effects may be caused by the defects in the magnesium plane [14]. In Ba(Fe$_{0.925}$Co$_{0.075}$)$_2$As$_2$, the calculated value of the intraband constant $\lambda_{22}^0 \leq 0.53$ is in good accordance with the value estimated from the optical experiments $\lambda_{22}^0 \approx 0.45$ [7].

**Conclusion**

We present a consistent theoretical analysis of general properties of two-gap superconductors performed within a standard theory based on generalization of a two-gap theory of superconductivity mediated by electron-phonon interaction. We show that, in contrast to regular BCS-superconductors, the large value of the ratio $2\Delta/T_c$ and of the electronic specific heat jump for the larger gap are no evidence for large electron-phonon interaction. We investigate the temperature dependence of the specific heat and the electron-phonon coupling constants for MgB$_2$ and Ba(Fe$_{0.925}$Co$_{0.075}$)$_2$As$_2$. The calculated temperature dependences of the electronic specific heat for the two compounds are in good accordance with experimental data. We show that the empirical two-band $\alpha$-model, frequently used to calculate the electronic specific heat, is applicable only to superconductors with relatively strong interband electron-phonon scattering.

Authors acknowledge the financial support of the Russian Foundation for Basic Research, grants 09-02-00560, 10-02-00614, 11-02-00199 and the Deutsche Forschungsgemeinschaft.

References.

[1] G. Gladstone, M.A. Jensen, J.R. Schrieffer, in *"Superconductivity"*, Ed. R.D. Parks, New York, 1969.

[2] R. A. Fisher, F. Bouquet, N. E. Phillips et al., Europhys. Lett. **56**, 856 (2001).

[3] F. Hardy, T. Wolf, R. A. Fisher et al., Phys. Rev. **B 81**, 060501(R) (2010)

[4] J.A. Liu, I.I. Mazin, J. Kortus, *Phys.Rev.Lett*. **87**, 087005 (2001)

[5] H. Padamsee, J. E Neighbor, and C. A. Shiffman, Journal of Low Temperature Physics, **12**, Nos. 3/4, 387 (1973)

[6] A.E.Karakozov, E.G.Maksimov, Ya.G.Ponomarev. JETP Letters **91**, 24 (2010).

[7] E. G. Maksimov, A. E. Karakozov, B. P. Gorshunov et al., Phys. Rev. **B 83**, 140502(R) (2011)

[8] E. J. Nicol, J. P. Carbotte, Phys. Rev. **B 71**, 054501 (2005)

[9] A.A.Abrikosov, L.P. Gor'kov, I.E. Dzyaloshinski *Methods of Quantum Field Theory in Statistical Physics* Dover, New York 1977.

[10] Ginzburg V.L., Kirzhnits D.A. (Eds) *High – Temperature Superconductivity* (New York: Consultants Bureau, 1982).

[11] A.J. Leggett, *Prog.Theor.Phys*. **36**, 901 (1966)

[12] Ponomarev Ya.G., Kuzmichev S.A., Mikheev M.G. et al., Solid State Comm. **129**, 85 (2004)

[13] B.Gorshunov, D.Wu, A.A.Voronkov et al., Phys. Rev. **B 81**, 060509(R) (2010)

[14] L. Boeri, O.V. Dolgov, A.A.Golubov. Phys. Rev. Lett. **101**, 026403 (2008)

[15] Jens Kortus, Oleg V. Dolgov, and Reinhard K. Kremer Phys. Rev. Lett **94,** 027002 (2005).

Table I. Parameters of the superconducting state calculated for a superconductor in the weak coupling regime (1 and 2), a layered superconductor of the family $Mg_{1-x}Al_xB_2$ with $T_c = 32K$ [6] ($MgB_2$), and an example that demonstrates the dependence of $\alpha_1$ (and $T_c$) on interaction $\tilde{\lambda}_{12}$ (labeled "$MgB_2$").

| sample | $\lambda^0_{11}$ | $\lambda^0_{22}$ | $\lambda^0_{12}$ | $\lambda^0_{21}$ | $\mu^*$ | $\theta(0)$ | $\tilde{\lambda}_{11}$ | $\tilde{\lambda}_{22}$ | $\tilde{\lambda}_{12}$ | $\tilde{\lambda}_{21}$ | $\alpha_1$ | $\alpha_1^{exp}$ |
|---|---|---|---|---|---|---|---|---|---|---|---|---|
| 1 | 0.4 | 0.2 | 0.2 | 0.15 | 0.1 | 0.29 | 16.2 | 6.4 | 5.4 | 3.2 | 2.04 | 2.04 |
| 2 | 0.4 | 0.12 | 0.2 | 0.15 | 0.1 | 0.29 | 13.9 | 5.2 | 0.93 | 2.62 | 1.82 | 1.82 |
| $MgB_2$ | 0.82 | 0.43 | 0.124 | 0.157 | 0.12 | 0.29 | 5.12 | 2.78 | 0.03 | 0.33 | 1.78 | 3.08 |
| "$MgB_2$" | – | – | – | – | – | 0.29 | – | – | 5.107 | 0.33 | 3.08 | – |

Table II. Gap values and coupling and other parameters characterizing superconducting Mg$^{11}$B$_2$ [2] and Ba(Fe,Co)$_2$As$_2$ [3].

| Sample /Reference | fit | $\tilde{\lambda}_{12}^{max}$ | $\tilde{\lambda}_{21}^{min}$ | $\alpha_1(\tilde{\lambda}_{12}^{max}, \tilde{\lambda}_{21}^{min})$ | $\alpha_2(\tilde{\lambda}_{12}^{max}, \tilde{\lambda}_{21}^{min})$ | $\Delta_1$ meV | $\Delta_2$ meV |
|---|---|---|---|---|---|---|---|
| Mg$^{11}$B$_2$ [2] | $\alpha'$ | 1.9 | 2.75 | 2.3088 | 0.5997 | 7.77 | 2.0 |
| | two-band $\alpha$ | – | – | 2.2 | 0.6 | 7.337 | 2.0 |
| Ba(Fe,Co)$_2$As$_2$ [3] | $\alpha'$ | 0.45 | 0.25 | 2.4875 | 0.9748 | 4.587 | 1.797 |
| | two-band $\alpha$ | – | – | 2.2 | 0.95 | 4.057 | 1.752 |

Table III. Electron-phonon interaction constants of superconductors $MgB_2$ [6], $Mg^{11}B_2$ [2] and Ba(FeCo)As [3]. For comparison, the data obtained for $MgB_2$ ($T_c$ =40.5 K) determined from the tunnel experiments results [6] is also presented.

| Sample/ Reference | fit | $\lambda_{11}^0$ | $\lambda_{22}^0$ | $\lambda_{12}^0$ | $\lambda_{21}^0$ | $\mu^*$ | $\Omega_{\tilde{n}}$ (meV) | $\theta(0)$ | $T_c$ (K) |
|---|---|---|---|---|---|---|---|---|---|
| $MgB_2$ [6] | | 0.921 | 0.43 | 0.124 | 0.157 | 0.12 | 67.76 | 0.25 | 40.5 |
| $Mg^{11}B_2$ [2] | max-min | 0.78 | 0.228 | 0.205 | 0.209 | 0.12 | 67.76 | 0.26 | 38.7 |
| | min-max | 0.767 | 0.204 | 0.124 | 0.22 | | | | |
| Ba(FeCo)As [3] | max-min | 0.889 | 0.531 | 0.204 | 0.157 | 0.12 | ≈30.0 | 0.39 | 21.4 |
| | min-max | 0.85 | 0.289 | 0.128 | 0.275 | | | | |

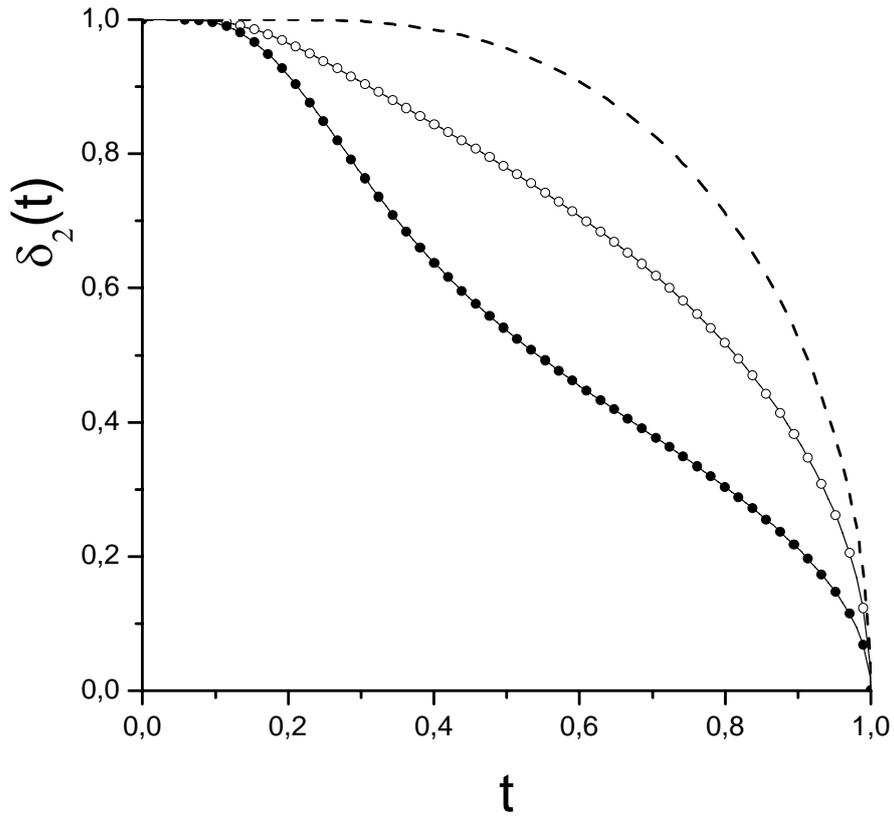

Fig.1. Dependence of reduced gaps $\delta_2(t) = \Delta_2(t)/\Delta_2(0)$ on the reduced temperature t = T/T$_c$ calculated with $\theta(0)$=0.275, $\alpha_1$=1.78 for two superconductors with different interband couplings, (a) $\tilde{\lambda}_{12}$=0.05, $\tilde{\lambda}_{21}$=0.1 (filled symbols) and (b) $\tilde{\lambda}_{12}$=0.22, $\tilde{\lambda}_{21}$=0.8 (open symbols). Dashed line shows the $\delta_1(t)$ dependence.

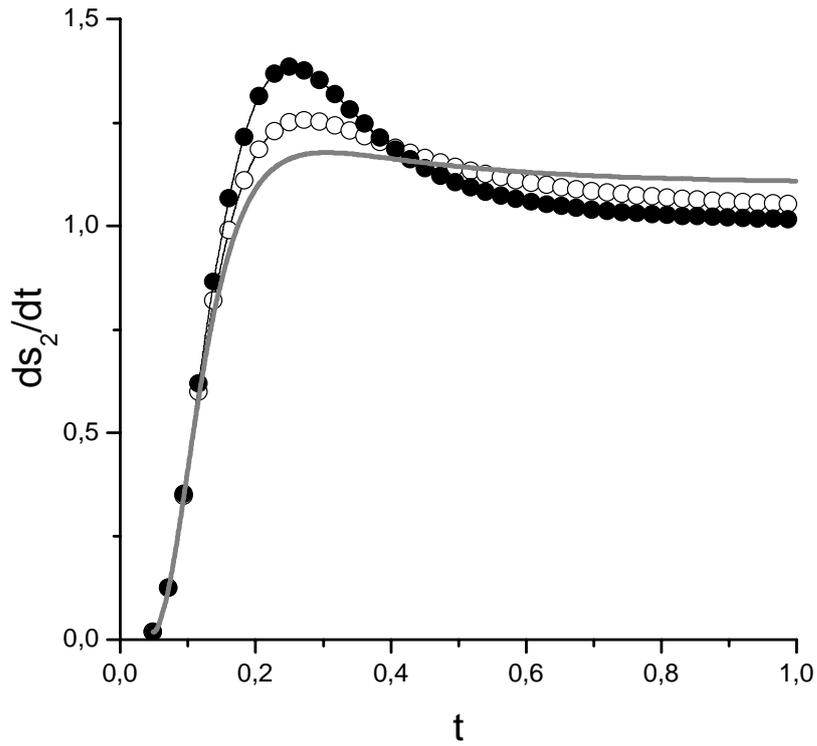

Fig. 2. Temperature dependence of normalized electronic specific heat of the band with smaller gap calculated with $\theta(0)=0.275$, $\alpha_1=1.78$ for two superconductors with different interband coupling, (a) $\tilde{\lambda}_{12}=0.05$, $\tilde{\lambda}_{21}=0.1$ (filled symbols) and (b) $\tilde{\lambda}_{12}=0.22$, $\tilde{\lambda}_{21}=0.8$ (open symbols). Solid line shows calculation according to the two-band $\alpha$-model.

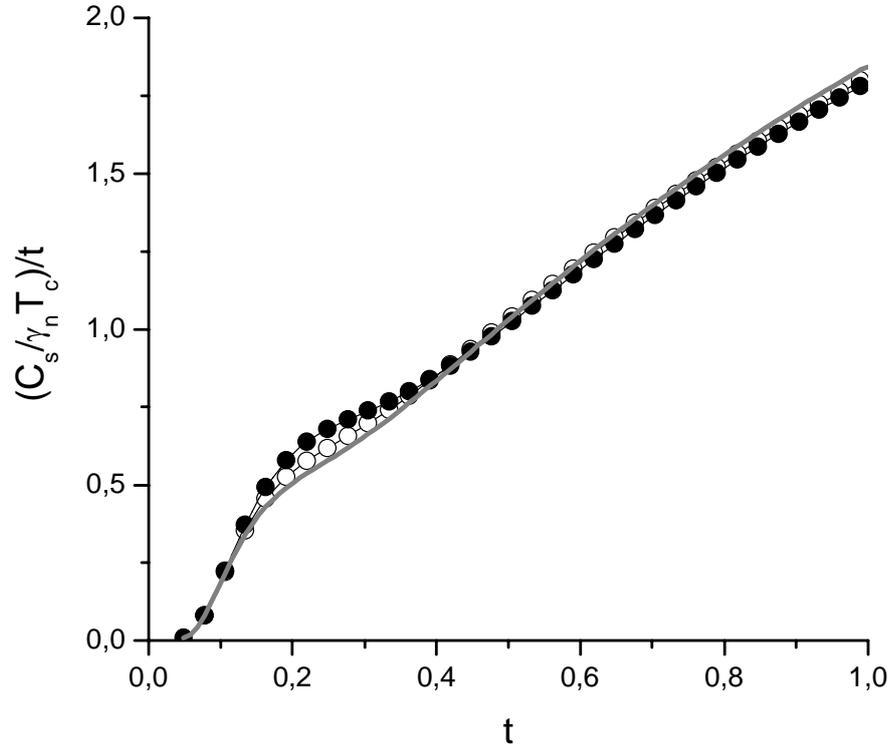

Fig. 3. Temperature dependence (t = T/$T_c$) of normalized electronic specific heat calculated with $\theta(0)=0.275$, $\alpha_1=1.78$ for two superconductors with different interband couplings, (a) $\tilde{\lambda}_{12}=0.05$, $\tilde{\lambda}_{21}=0.1$ (filled symbols) and (b) $\tilde{\lambda}_{12}=0.22$, $\tilde{\lambda}_{21}=0.8$ (open symbols). Solid line corresponds to calculations according to the two-band $\alpha$-model with $\gamma_1/\gamma_n=0.55$ and $\gamma_2/\gamma_n=0.45$.

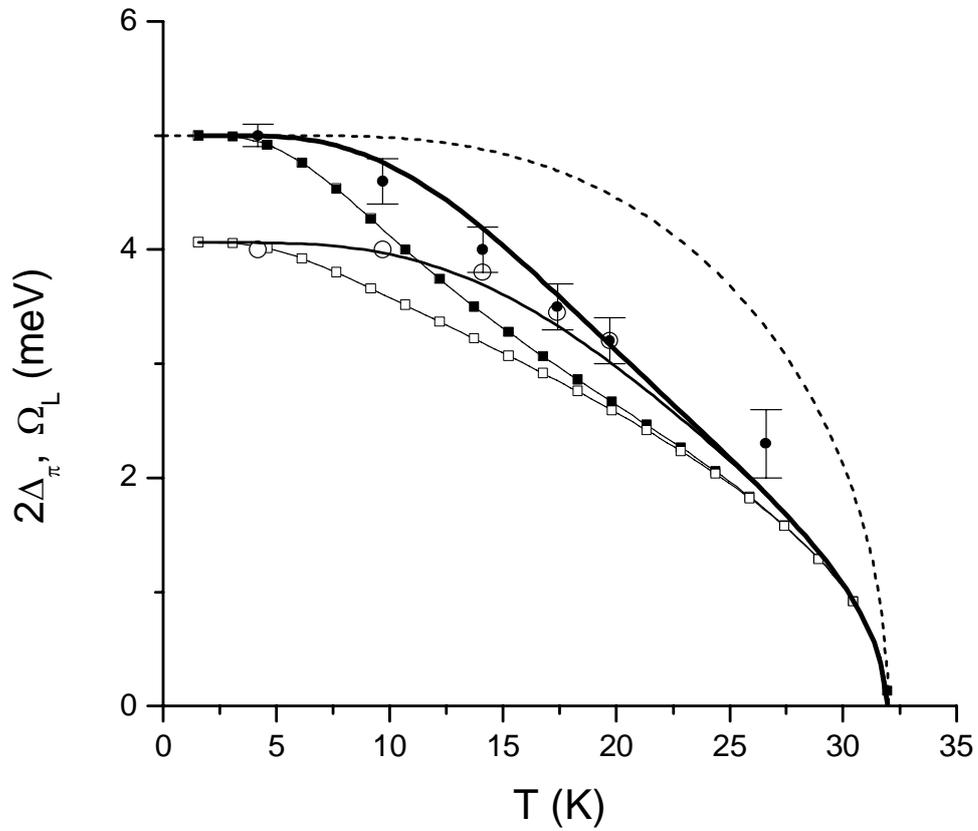

Fig. 4. Temperature dependence of the gap $2\Delta_\pi$ (filled symbols) and of the Leggett mode energy $\Omega_L$ (empty symbols) of $Mg_{1-x}Al_xB_2$ with $T_c = 32K$ [12]. Thick and thin solid lines – calculations within the $\alpha'$-approximation of the gap and Legget mode energies, respectively. Lines with filled and empty square symbols – calculation according to the $\alpha$-approximation of the of the gap and Legget mode energies, respectively.

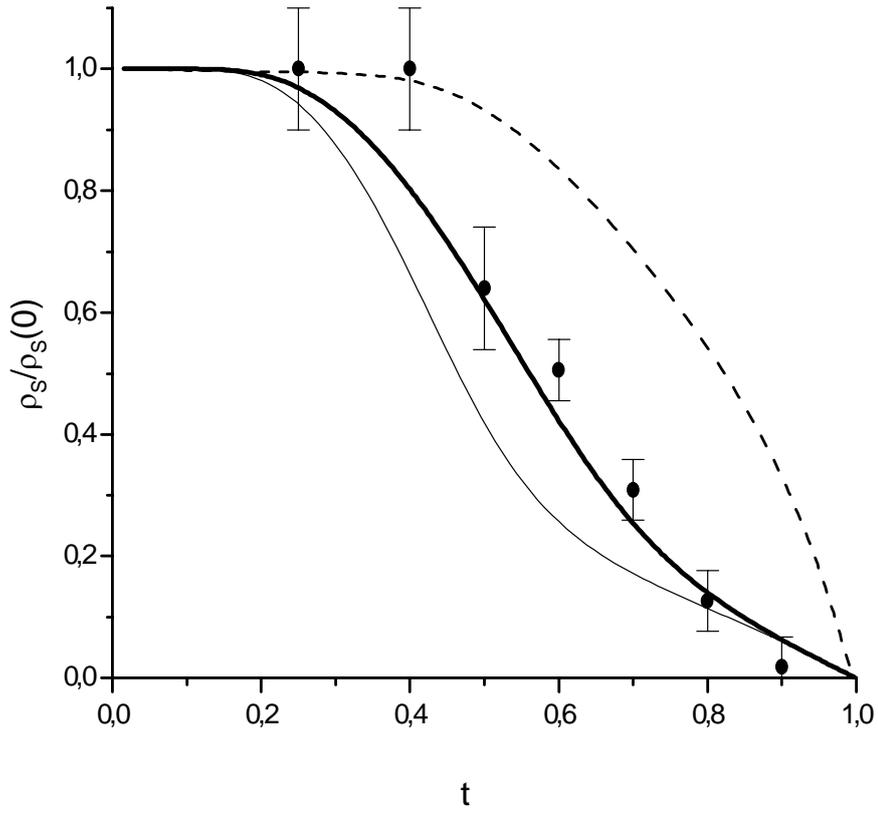

Fig. 5. Temperature dependence of normalized superconducting condensate density of Ba(Fe$_{0.9}$Co$_{0.1}$)$_2$As$_2$ [13] (symbols). Lines correspond to calculations according to the $\alpha'$-approximation (thick), $\alpha$-approximation (thin) and two-band $\alpha$-model with $\tilde{\lambda}_{12} \sim \tilde{\lambda}_{21} \sim 0.1$ (dashed).

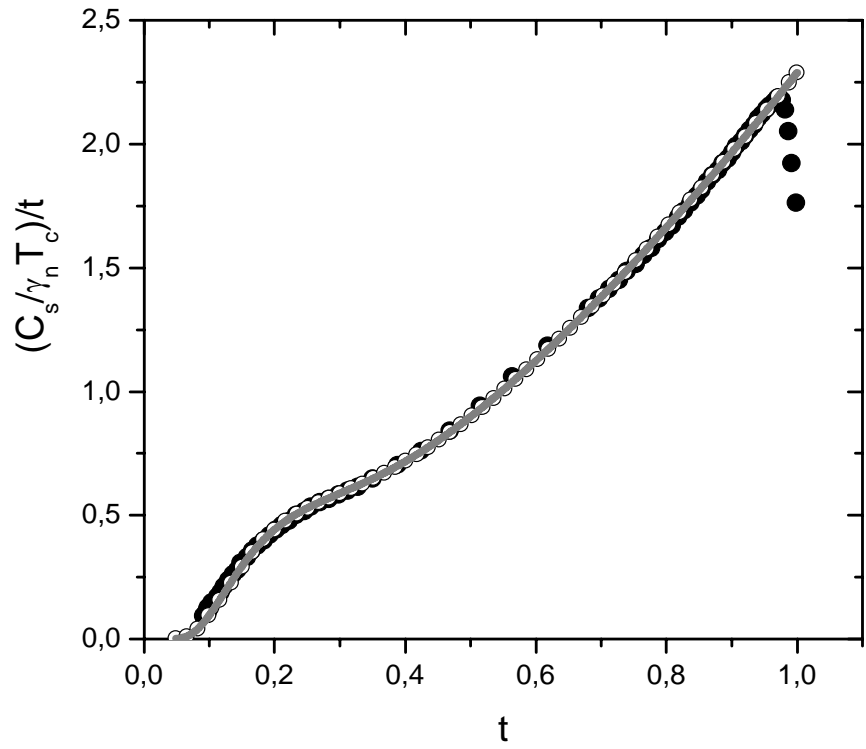

Fig. 6. Temperature dependence of electronic specific heat of $Mg^{11}B_2$ [2] (filled symbols). Lines show calculations according to the two-band $\alpha$ – model (grey line) and $\alpha'$ – approximation with parameters given in Table II (open symbols).

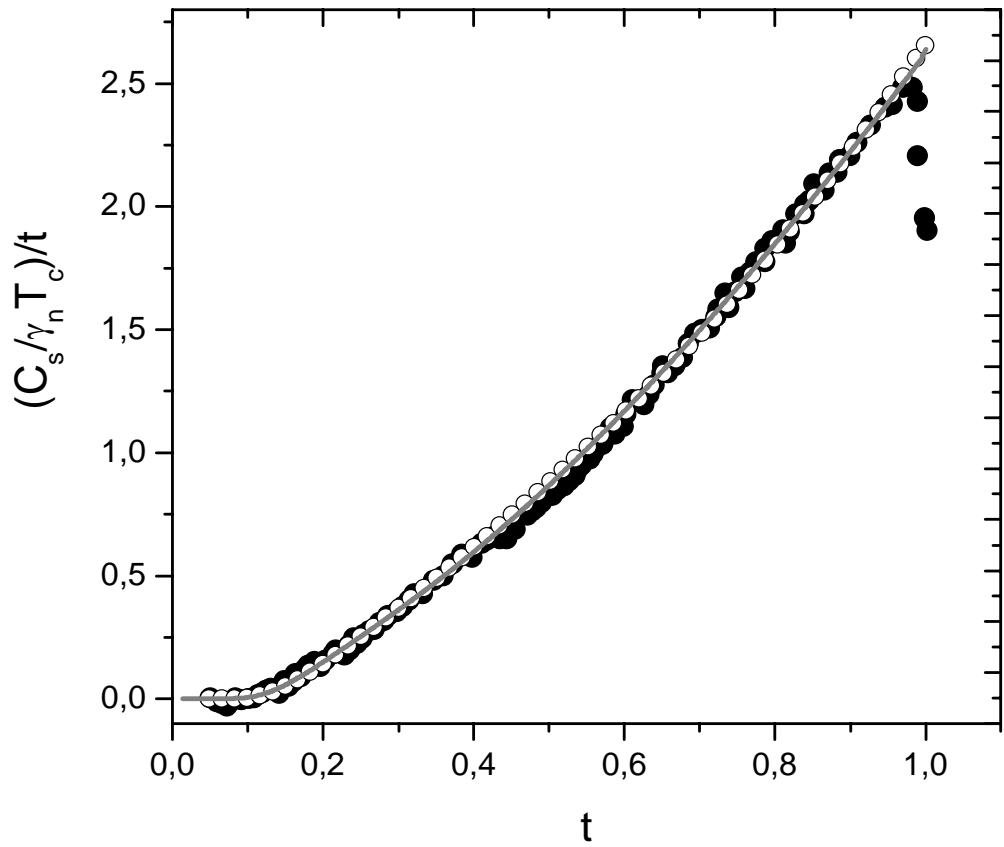

Fig. 7. Temperature dependence of electronic specific heat of Ba(Fe$_{0.925}$Co$_{0.075}$)$_2$As$_2$ [3] (filled symbols). Lines show calculations according to the two-band $\alpha$ – model (grey line) and $\alpha'$ – approximation with parameters given in Table II (open symbols).